\documentclass[pre,twocolumn,aps]{revtex4}
\usepackage{times}
\usepackage{amsmath}
\usepackage{graphicx}
\usepackage{amssymb}
\usepackage{color}
\usepackage{amsfonts}

\newcommand{\ii}{\mathrm{i}}
\newcommand{\wb}{\omega_\mathrm{b}}
\newcommand{\wo}{\omega_0}

\begin{document}

\title{Floquet Analysis of Kuznetsov--Ma breathers: A Path Towards %the
Spectral Stability of Rogue Waves}

\author{J. Cuevas-Maraver}
%\email[Email: ]{jcuevas@us.es}
\affiliation{Grupo de F\'{i}sica No Lineal, Departamento de F\'{i}sica Aplicada I,
Universidad de Sevilla. Escuela Polit\'{e}cnica Superior, C/ Virgen de \'{A}frica, 7, 41011-Sevilla, Spain \\
Instituto de Matem\'{a}ticas de la Universidad de Sevilla (IMUS). Edificio Celestino Mutis. Avda. Reina Mercedes s/n, 41012-Sevilla, Spain}

\author{P. G. Kevrekidis}
%\email[Email: ]{kevrekid@math.umass.edu}
\affiliation{Department of Mathematics and Statistics, University of Massachusetts
Amherst, Amherst, MA 01003-4515, USA}

\author{D. J. Frantzeskakis}
%\email[Email: ]{dfrantz@phys.uoa.gr}
\affiliation{Department of Physics, National and Kapodistrian University of Athens, Panepistimiopolis,
Zografos, Athens 15784, Greece}

\author{N. I. Karachalios}
\affiliation{Department of Mathematics, University of the Aegean, Karlovassi, 83200 Samos, Greece}

\author{M. Haragus}

\affiliation{Institut FEMTO-ST, D\'{e}partement OPTIQUE, Univ. Bourgogne Franche-Comt\'{e},  25030 Besan\c{c}on Cedex, France}

\author{G. James}
\affiliation{INRIA Grenoble - Rh\^one-Alpes, Bipop Team-Project, Inovall\'ee, 655 Avenue de l'Europe, 38334 Saint Ismier Cedex, France.}

%\date{\today}

\begin{abstract}
In the present work, we aim at taking a  step towards the spectral stability analysis of Peregrine solitons, i.e., wave structures that are used to emulate extreme wave events. Given the space-time localized nature of Peregrine solitons, this is a priori a non-trivial task. Our main tool in this effort will be the study of the spectral stability of the periodic generalization of the Peregrine soliton in the evolution variable, namely the Kuznetsov--Ma breather. Given the periodic structure of the latter, we compute the corresponding Floquet multipliers, and examine them in the limit where the period of the orbit tends to infinity. This way, we extrapolate towards the stability of the limiting structure, namely the Peregrine soliton.   We find that multiple unstable modes of the background are {\it enhanced}, yet no additional unstable eigenmodes arise as the Peregrine limit is approached. We explore the instability evolution also in direct numerical simulations.
\end{abstract}

\maketitle

\section{Introduction}

The study of extreme events and
rogue (or freak) waves is a topic
that has gained tremendous momentum during the last decade~\cite{k2a,k2b,k2c,k2d}.
Part of the appeal of the subject lies in the emergence
of a diverse array of relevant experimental observations
in systems ranging
from hydrodynamics~\cite{hydro,hydro2,hydro3}
to superfluid helium~\cite{He}, from nonlinear optics~\cite{opt1,opt2,opt3,opt4,opt5,laser}
to plasmas~\cite{plasma}, and from Faraday surface ripples
\cite{fsr} to parametrically driven capillary waves \cite{cap}.
On the theoretical side,
candidate structures related to extreme events
in prototypical, nonlinear Schr{\"o}dinger (NLS) type models,
have emerged from the seminal works of Peregrine~\cite{H_Peregrine}, Kuznetsov~\cite{kuz}, Ma~\cite{ma},
and Akhmediev~\cite{akh}, as well as Dysthe and Trulsen~\cite{dt}.
In turn, the existence of relevant structures and their wide
experimental realizability,
has motivated a broad array of
theoretical works, now summarized in many reviews
\cite{yan_rev,solli2,onorato}.

Perhaps the most notable example among the theoretically analyzed
solutions in the form of a rogue wave is the so-called Peregrine
soliton (PS)~\cite{H_Peregrine}. Its algebraic decay in both space
and time render it a natural candidate for a wave that
``appears out of nowhere and disappears without a trace''~\cite{akmpla}.
Given its significance and emergence in a
broad array of physical settings, a natural question that
arises for the PS
is that of its stability. To examine the robustness of the PS,
various approaches have been considered.
One of them is to explore to what extent the existence of such
a rogue wave will be preserved under perturbations;
in that context, it was found~\cite{devine} that the PS
can still be identified in Hirota-type variants of the original NLS equation (see
also Ref.~\cite{calinibook} for relevant work in higher-order NLS models).

Another approach for the study of robustness of extreme waves relies on
the connection with the so-called Kuznetsov-Ma breather
(KMb) \cite{kuz,ma}. This latter state is periodic in the evolution
variable and, in the limit of infinite period, reduces to the PS.
The dynamical robustness of the KMb against dispersive or diffusive (dissipative) perturbations
was studied by means of an adiabatic approximation \cite{cagnon} and, more recently,
by means of a perturbed inverse scattering transform approach \cite{KalimerisGarnier}.
In the latter work it was shown that the KMb is rather robust against dispersive perturbations,
but it is strongly affected by dissipative ones. Notice that direct numerical simulations
with perturbed initial data have also been used to explore the robustness of the PS
\cite{marianna} and the KMb \cite{calini14}.

On the other hand, the modulational instability (MI) of the background
plane wave (which hosts the KMb or the PS) leads to homoclinic-type solutions,
which can also be considered as candidates for rogue waves \cite{calinibook,calini2012}.
The study of the criteria for the emergence and stability of such homoclinic-type solutions
revealed that spatial phases of a multi-mode homoclinic orbit can be optimally selected,
so that the modes coalesce at a given time, leading to significant wave amplification
beyond that predicted by the typical MI. This approach has been effectively applied to
a variety of extended NLS models involving higher-order dissipative or dispersive effects,
including the Dysthe model \cite{dysthe}.

Another possibility, more proximal to the spirit of our considerations,
was that of considering the spectral stability analysis around a PS
\cite{vangorder}; for a similar attempt both
around the PS and around the KMb, see the works of~\cite{abdul1,abdul2}.
However,
in that case, the fundamental complication
consists of assigning a meaning to the linearization around
a time-dependent state. While in other problems, especially dissipative
ones~\cite{wayne},
a ``frozen time'' spectrum and its usefulness
have been made more precise, in our setting this is not sufficiently
clear, to the best of our understanding; hence, we will not follow %this
such an approach here.

In the present work, we will instead focus on a variant of the
problem (of the PS stability) for which the notion of
a spectrum is well-defined. In particular, we will consider the KMb which,
due to its periodicity in the evolution variable,
is amenable to spectral analysis by means
of Floquet methods. At the same time,
the KMb solution encompasses
the case of the PS solution in the special limit of infinite period, as mentioned above.
Our plan is then a natural one: we will consider the Floquet analysis of the KMb, varying
the period, in an effort to appreciate the trend towards
the limiting case of the PS. By following the relevant multipliers
as the KMb gradually morphs into the PS (for a visually
appealing demonstration of this process, see, e.g.,~\cite{youtube}),
we expect to acquire a sense of the relevant limit case as far as
spectral stability is concerned.
We find, as can arguably be anticipated, that the KMb carries
instability modes associated to the MI of its non-vanishing background.
Intriguingly, the growth rates of a number of the relevant modes are found to be higher
than those in the absence of the PS. However, we also observe that no additional instabilities arise
on top of the background instabilities. We illustrate
results of detailed numerical continuations, as well as explore the dynamical evolution of the
associated instabilities in direct numerical simulations.

The paper is structured as follows. In Section~II, we present the model, the
KMb and PS solutions and report results of our numerical simulations; these include
Floquet spectral analysis and evolution simulations. Finally,
in Section III, we summarize and discuss the implications of our results
with an eye towards future work.
\section{Numerical Results}
\label{SecA}
\subsection{Theoretical and Numerical Setup}
We consider the stability of the KMbs in the prototypical
setting of the focusing NLS
equation~\cite{sulem,ablowitz}:
\begin{equation}\label{eq:nls}
    \ii \partial_t u+\frac{1}{2}\partial_{x}^2 u+(|u|^2 -1) u=0.
\end{equation}
Here, $u$ denotes the complex order parameter (which
%can
represents, e.g., the electric field envelope in optics),
subscripts denote partial derivatives, while it should be noted that
the evolution variable in this setting is $t$ (representing the propagation distance in optics).

The NLS Eq.~(\ref{eq:nls})
possesses the exact analytical KMb solution, which is of the following form:
\begin{equation}
    \label{eq:kmb}
    u_{\mbox{\tiny KM}}(x,t)=1+\frac{2(1-2a)\cos(\wb t)-\ii\wb\sin(\wb t)}{\sqrt{2a}\cosh(bx)-\cos(\wb t)},
\end{equation}
where $a=(1+\sqrt{\wb^2 +1})/4$ and $b=2\sqrt{2a-1}$, {for temporal frequency $\wb>0$.
In the limit
$\wb\rightarrow 0$ (corresponding to temporal period $T=2\pi/\wb \rightarrow \infty$),
i.e., for $a=1/2$, the KMb reduces
to the PS
which is given by:}
\begin{equation}
u_{\mbox{\tiny PS}}(x,t)=1-\frac{4 (1+2 \ii  t)}{1+4  x^{2}
+4  t^2}.
\label{par_peregrine}
\end{equation}
{Notice that} the NLS Eq.~(\ref{eq:nls})
admits a Galilean invariance, i.e.,
%is Galilei invariant, i.e.,
each solution $u(x,t)$ provides the
one-parameter family of solutions
\begin{equation}\label{eq:gali}
\tilde{u}_c(x,t)=e^{i\, (c\, x-\frac{c^2}{2}\, t)}\, u(x-c\, t,t),
\end{equation}
where $c$ is an arbitrary constant. Consequently, Eq.~(\ref{eq:nls}) also admits traveling
KMb and PS solutions.

{Linear stability} of the KMb can be analyzed by means of Floquet analysis --cf. details
in Appendix A. To this effect, the time evolution of a small perturbation $\xi(x,t)$ to a
given {time-periodic solution, in our case the KMb,} must be traced. This
perturbation is introduced in Eq.~(\ref{eq:nls}) as \[u(x,t)=u_{\mbox{\tiny KM}}(x,t)+\delta
\xi(x,t),\] where $\delta$ is a formal small parameter. The
resulting equation at order $O(\delta)$ reads: %as
\begin{equation}\label{eq:perturb}
  \left( i \partial_t+\frac{1}{2}\partial_{x}^2+2|u_{\mbox{\tiny KM}}|^2-1\right)\xi
  + u_{\mbox{\tiny KM}}^2 \bar{\xi}=0,
  %^{\star}=0,
\end{equation}
where $\bar{\xi}$ denotes the complex  conjugate of $\xi$.

In the framework of Floquet analysis, the stability properties of periodic
solutions are determined by the eigenvalues of the monodromy matrix, $\mathcal{M}$,
which is defined as:
\begin{equation}
\label{MM}
\left[\begin{array}{c}
  \mathrm{Re}(\xi(x, T)) \\ \mathrm{Im}(\xi(x, T)) \\  \end{array}\right]
  =\mathcal{M}
  \left[\begin{array}{c}
  \mathrm{Re}(\xi(x, 0)) \\ \mathrm{Im}(\xi(x, 0)) \\  \end{array}\right] ,
\end{equation}
where it is reminded that $T=2\pi/\wb$. Due to the Hamiltonian structure
of the system, linearly stable periodic solutions are neutrally stable.
More precisely, in the neutral stability case, all the eigenvalues $\lambda$
of the monodromy matrix $\mathcal{M}$ [also called Floquet multipliers (FMs)],
lie on the unit circle. Furthermore, for such systems, the set of Floquet
exponents $\sigma$ satisfying $\lambda=\exp(\sigma T)$ is symmetric with
respect to both the real, and imaginary axis, in the complex plane. Consequently,
the set of Floquet multipliers (the Floquet spectrum) is symmetric with respect
to both transformations $\lambda\mapsto\bar\lambda$ and $\lambda\mapsto1/\lambda$.
Therefore, when instabilities arise, they can be of two types:
\begin{itemize}
\item[(a)] due to eigenvalues coming out of the unit circle (at $\lambda=1$)
in real pairs $(\lambda,1/\lambda)$, in which case the instability is {\it exponential}
(the relevant {Floquet exponent $\sigma$ is real} and
the growth is purely exponential); %or
\item[(b)] {due {to} eigenvalues coming}
out in complex quartets, in which case
the instability possesses both a growing and
an oscillatory part and, hence, is referred to as an {\it oscillatory} one.
\end{itemize}

In order to numerically implement the monodromy calculation, one must
integrate the perturbation equation (\ref{eq:perturb}).
As the KMb solution possesses a relatively steep peak, this equation is mildly stiff.
Therefore, we use, as a suitable option for
the numerical integration,
the Exponential Time Differencing 4th-order
Runge--Kutta method (ETDRK4), with Fourier spectral collocation;
this implies periodic boundary conditions in the domain $[-L,L]$
\cite{etd}.
%{\bf
This method, upon splitting the problem into a (potentially stiff)
linear and (likely non-stiff) nonlinear part uses the Duhamel formula
and an RK4  method to approximate the relevant resulting time integral.
Details of the method (originating from the work of~\cite{cox}),
including issues of stabilization and relevant de-aliasing are discussed
in an intuitive way in~\cite{kassam2}.
%}

We have chosen a discretization parameter $\Delta x=0.05$ and a time step $\Delta t=(\Delta x)^2/4$,
which ensures the stability of the numerical integrator. Below we present results
for domains with $L=5$ and $L=10$. We also compare these with the findings
for $L=15$ in order to showcase the trend of the relevant stability
results as the domain size is increased.

\subsection{Stability Analysis}

{{As it is evident from Eq.~(\ref{eq:kmb}),} KMbs are time-periodic solutions with
a localized spatial structure. {In particular,} the analytical expression in
Eq.~(\ref{eq:kmb}) shows that as the spatial variable $|x|\to\infty$, a KMb solution
exponentially converges towards the constant (representing a plane wave) $u_\infty=1$.
{The latter} is also the asymptotic state of the PS. By analogy to the case of
localized steady, or traveling waves, we therefore expect that the Floquet spectrum
of a KMb consists of an
``essential'' spectrum, which is the same as the one of
the asymptotic state $u_\infty=1$, and possibly some additional isolated eigenvalues
popping out of this spectrum or associated with symmetries of the solution.
In particular, instabilities of $u_\infty=1$ should induce instabilities of KMbs.}

The plane wave background $u_\infty=1$ is known to be modulationally
unstable~\cite{zakharov}. The frequency $\wo$ and wavenumber $k$ of a perturbation
around $u_\infty=1$, as obtained from a standard modulational instability (MI)
analysis, are connected through the dispersion relation:
%is given by
\begin{equation}
\label{eq:wo}
\wo = \pm \frac{1}{2}k\, (k^2 -4)^{1/2},
\end{equation}
%at wavenumber $k$.
The MI {analysis} on the real line {gives rise to an} instability band,
namely a band of unstable frequencies, for wavenumbers $k \in [-2,2]$.
The frequency $\wo$ is related to the FMs as:
%according to
\begin{equation}
\label{approxlambda}
\lambda = \exp(2\ii\pi\wo/\wb).
\end{equation}
It turns out from Eq.~(\ref{approxlambda}), that the Floquet spectrum of
$u_\infty=1$ consists of the unit circle, together with a band of positive
Floquet multipliers arising in pairs $(\lambda,1/\lambda)$, due to wavenumbers $k \in [-2,2]$.

The Floquet analysis of KMbs numerically shows the existence of
double pairs {$(\lambda,1/\lambda)$ of positive FMs} different from $1$.
These FMs correspond to the MI of the plane wave background,
and are very close to the prediction from the MI analysis, taking into
regard the quantization of wavenumbers $k=\pi n/L$ ($n \in \mathbb{Z}$)
arising from the periodic boundary conditions.
The corresponding unstable eigenmodes
present a substantial ``hybridization'' with the breather.
By this notion, it is meant that the spatial profile of these eigenmodes
is not identical to the one of eigenmodes associated with the linearization around the background,
but rather carries also a local imprint at the center due to the presence of the breather.

%%%%%%%%%%%%%%%%%%%%%%%%%%%%%%%
\begin{figure}[tbp]
	\begin{center}
		\begin{tabular}{cc}
			\includegraphics[width=.34\textwidth]{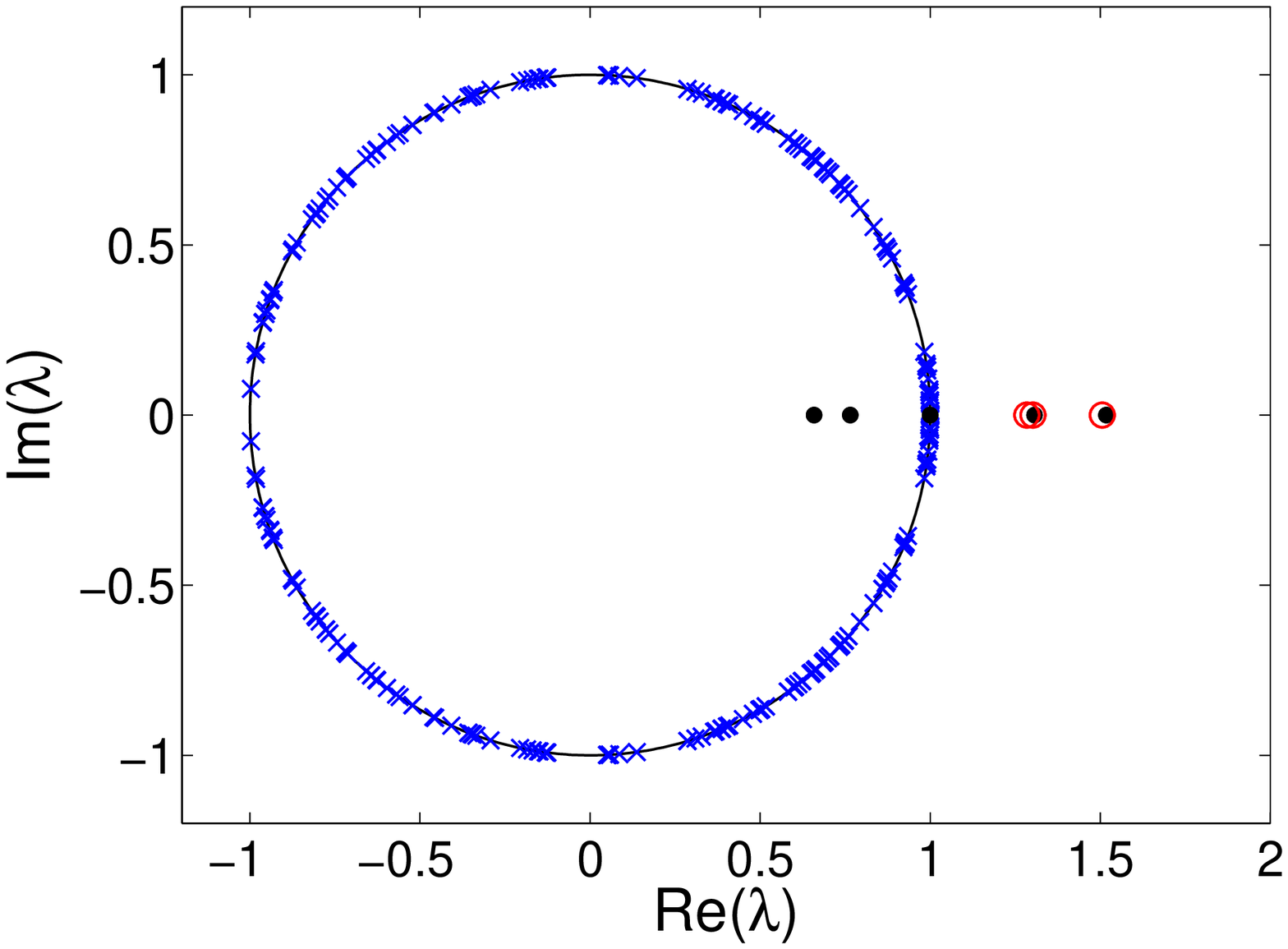} \\
			\includegraphics[width=.33\textwidth]{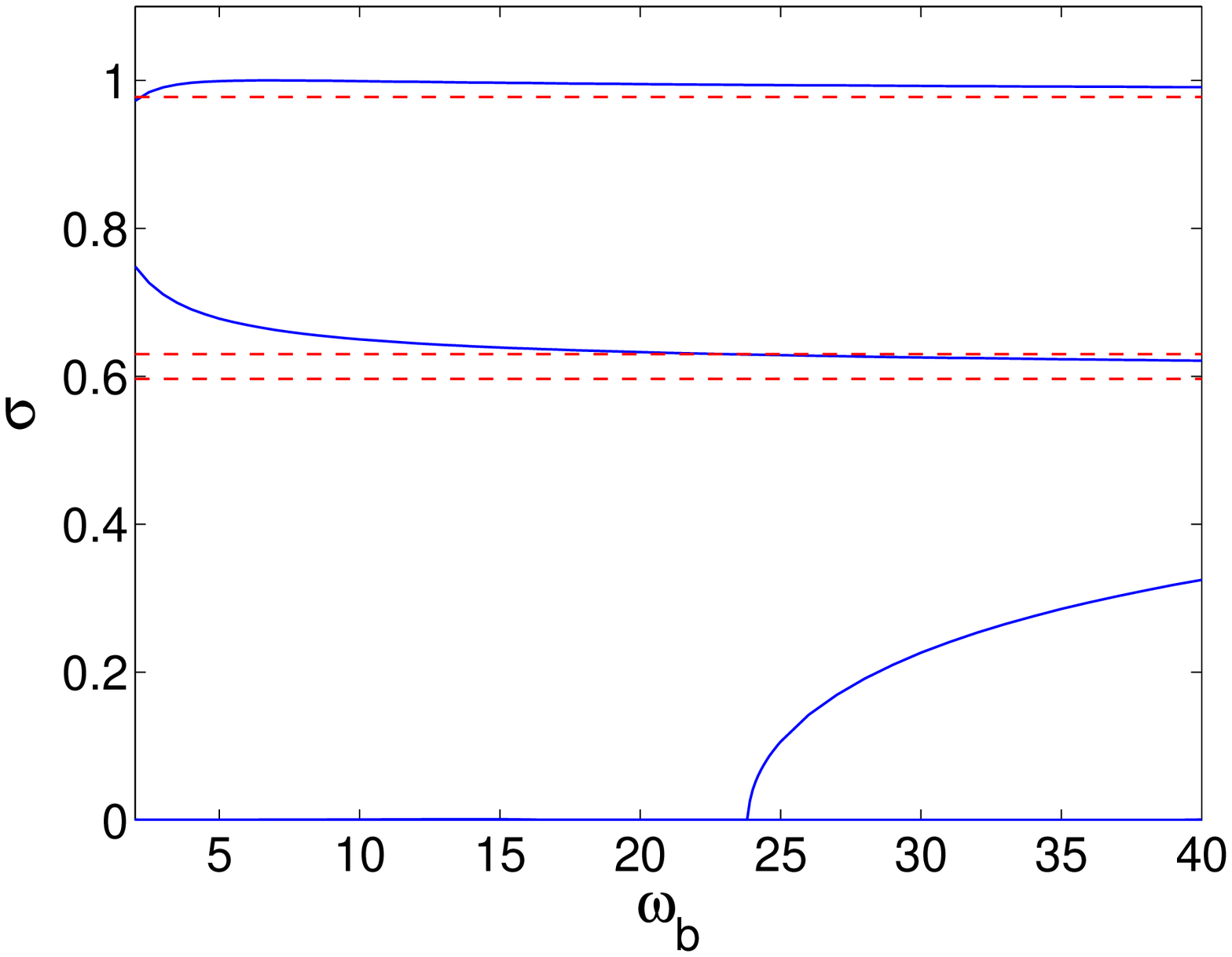}
		\end{tabular}
	\end{center}
	\caption{(Color online)
		Top panel: Floquet multiplier spectrum of the KMb
		for $L=5$ and $\wb=15$; open (red) circles correspond to the unstable
		eigenvalues, theoretically predicted
		via the MI analysis.
		The unstable and neutral modes of the linearization are presented
		with (black) dots, while the neutrally stable ones on the unit circle are given
		by (blue) $\times$'s.
		{Bottom panel: the dependence of the unstable eigenvalues $\sigma$}
		on the frequency $\omega_b$, for $L=5$.
		The solid line corresponds to the KMb numerically obtained multipliers,
		and dashed lines to the eigenvalues, once again theoretically predicted by the MI.}
	\label{fig1}
\end{figure}

Apart from these modes, there are three pairs at $\lambda=1$:
one of them corresponds to the translational mode, and another one
to the phase mode (due to the conservation of the $L^2$ norm, i.e., the power in %nonlinear
optics). The third pair stems, as is customary for breathers in Hamiltonian
systems, from the time translational invariance.
Finally, the remaining modes lie on the circle and correspond to background
modes with a rather weak hybridization with the breather.

The top panel of Fig.~\ref{fig1} shows a summary of the above results,
illustrating a typical example of the obtained FM spectrum and the instability
eigenmodes, in the case
$L=5$ and
$\wb=15$.
This case leads to {three} double unstable multipliers {for large $\wb$}
(as the domain expands, the number of unstable eigenmodes increases).
{One of these eigenvalues returns to the unit circle when $\wb\approx23.9$}.
The bottom panel depicts the dependence of the Floquet exponents $\sigma$
associated with the unstable modes, on the frequency of the breather.
Notice that in the limit of large $\wb$, the unstable Floquet
exponents approach their asymptotic values predicted by the MI
analysis above. On the other hand, for the FMs, $|\lambda|\rightarrow 1$,
as $\omega_b\rightarrow\infty$. This limiting behavior of the FM modulus
can be explained as follows: for $x\neq 0$, it follows from Eq.~(\ref{eq:kmb})
that in the limit $\wb\rightarrow\infty$, the solution $u_{\mbox{\tiny KM}}$
converges to $u_{\infty}=1$, of FM modulus $|\lambda|=1$.
However, the convergence of  $u_{\mbox{\tiny KM}}$ to the asymptotic
state $u_{\infty}=1$ is not uniform. For $x=0$, in the limit
$\omega_b\rightarrow\infty$, the solution  $u_{\mbox{\tiny KM}}$ diverges.
Hence, the KMb solution, being continuous itself as a function of the
parameter $\wb$, gives rise in the limit of large $\wb$, via a slow,
square-root rate divergence, to a ``solitonic excitation'' around $x=0$.
This solitonic excitation is of finite localization width
(for large, but finite $\wb$) and of increasing (to infinity) amplitude.
In accordance with Eq.~(\ref{eq:kmb}), the limiting solitonic excitation
on top of the unit background should consist of time-periodic peaks
of vanishing period $T=2\pi/\wb \rightarrow 0$, as $\wb\rightarrow\infty$.
This, however, does not affect the asymptotic behavior of the
FM modulus $|\lambda|\rightarrow 1$.

Of particular importance %to us
is the opposite limit where the KMb approaches the PS.
This is shown in the bottom panel of Fig.~\ref{fig1}.
It can be seen there that the lowest pair of the unstable modes
returns to the unit circle, when $\wb\approx 23.9$.
However, there are also modes like the second (blue solid curve)
mode in that panel whose growth rate monotonically increases
as $\omega_b$ decreases. This is suggestive that the structure
becomes progressively more unstable as the PS limit is approached
--although it should be clarified that our computations are never
able to precisely capture the limit; they can merely be characterized
as strongly suggestive of the approach to it.
There are also modes, like the one associated with the
top-most (blue solid) curve in the panel,
which may have a non-monotonic dependence on $\omega_b$, yet are
still larger, for all values of the breather frequency, than their
plane wave limit. Overall, this suggests that the presence of
the KMb (and by virtue of the limiting process, the PS)
enhances the instability of the background. Our results suggest
that this conclusion extends towards the PS state which
is approximated by the KMb as $\omega_b \rightarrow 0$.

To explore how our results depend on the size of the domain,
we repeat the computationally expensive FM computation in the case of a
domain twice as large, i.e., $L=10$ (and have confirmed the
conclusions also by extending considerations even to $L=15$).
In the case of $L=10$, we observe a similar scenario aside from the
feature that there now exist six double unstable multipliers for high frequency.
This is due to the fact that the increase of the domain size
results in a ``finer'' quantization of wavenumbers which, in turn,
means that more of the resulting wavenumbers find themselves in the
unstable band of the MI spectrum.
%%%%%%%%%%%%%%%%%%%%%%%%%%%%%%%%%%%%%
\begin{figure}[tbp]
	\begin{center}
		\begin{tabular}{cc}
			\includegraphics[width=.34\textwidth]{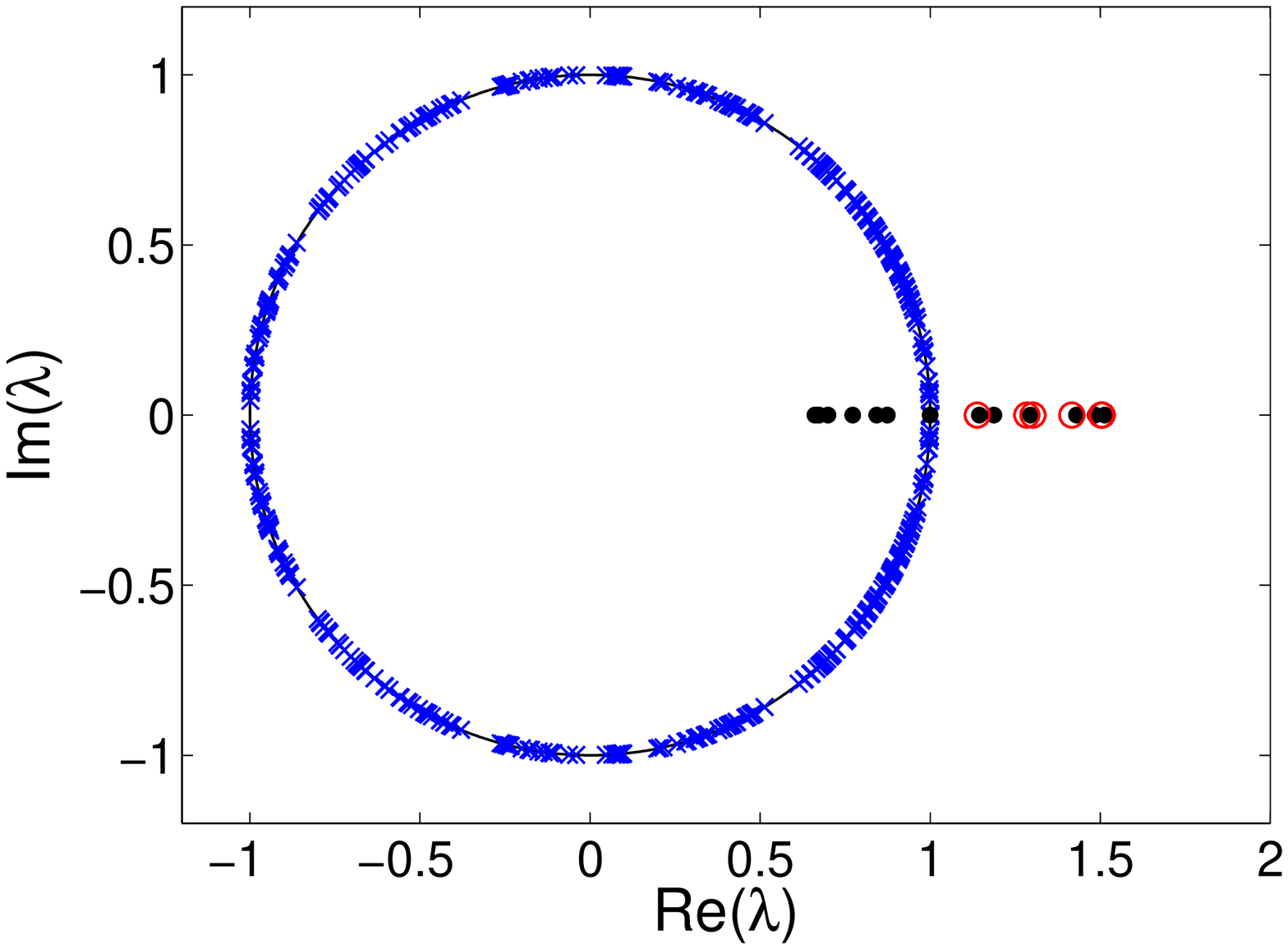} \\ %&
			\includegraphics[width=.33\textwidth]{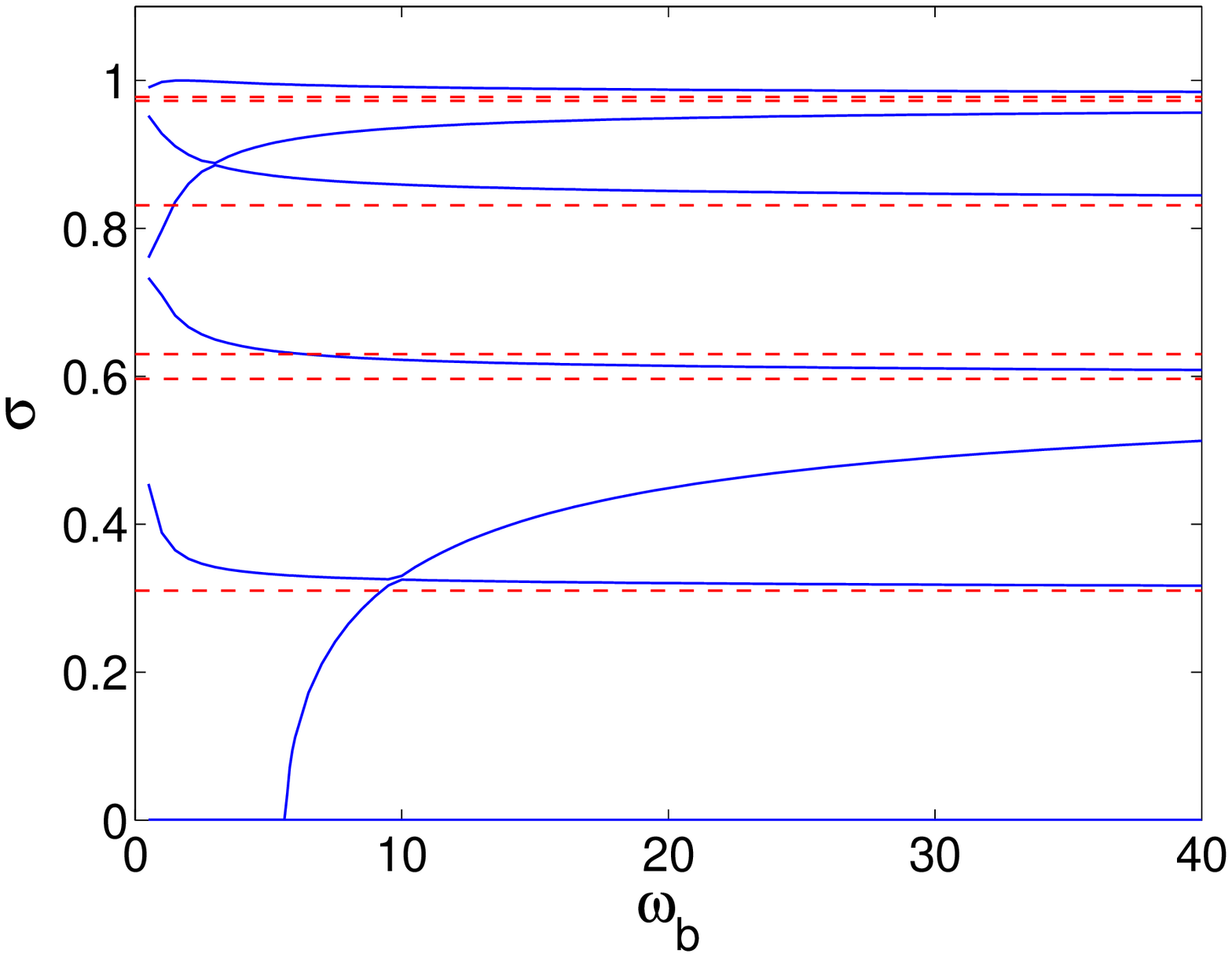} %\\
		\end{tabular}
	\end{center}
	\caption{(Color online) Same as Fig.~\ref{fig1}, but for $L=10$. It is clear that
		in this case the number of unstable modes is larger (in particularly,
		there are six, as can be counted in the
		bottom panel, shown by blue solid line, with their respective
                limiting values from the MI analysis given by red
                dashed line).}
	\label{fig2}
\end{figure}
%%%%%%%%%%%%%%%%%%%%%%%%%%%%%%%%%%%%%%%%%%%%%%%%%%%%%%%%%%%%%%%%%%%%

Figure~\ref{fig2} summarizes the aforementioned results for $L=10$,
similarly to Fig.~\ref{fig1}. A key feature becomes evident
in both of the bottom panels of Figs.~\ref{fig1} and \ref{fig2}:
there exist multiple unstable modes that have a stronger instability
growth rate {\it in the presence} of the localized pattern (i.e., the KMb),
rather than in its absence. This is true overall even though one pair of modes
disappears into the unit circle for $\omega_b \approx 5.7$.
Most of the rest of them (e.g., three clearly discernible
examples in the bottom of Fig.~\ref{fig2}) increase for
decreasing $\omega_b$, while a couple may have non-monotonic
or even decreasing dependence over $\omega_b$, although the
overall growth rate is always higher in the presence of
the KMb in comparison to the limit predicted by MI of the background
state (dashed lines in the relevant panel).
This is an interesting finding, indicating that the
instability of the background is, in fact, overall enhanced
by the presence of the KMb, and is strongly suggestive that
this feature persists as the frequency is decreased
towards the PS limit.
On the other hand, that being said, it should be highlighted
that there are no new unstable modes emerging solely due to
the presence of the KMb: specifically, there is {\it no point spectrum}
(no localized eigendirections) associated with the instability of the KMb.

\begin{figure}[tbp]
	\begin{center}
		\begin{tabular}{cc}
			\includegraphics[width=.34\textwidth]{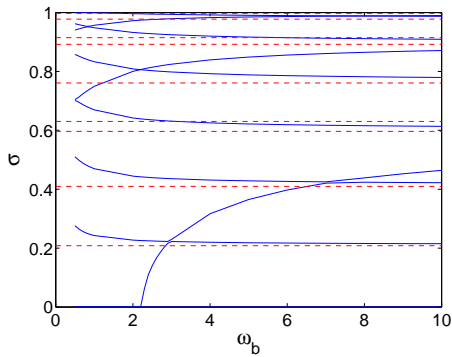}
		\end{tabular}
	\end{center}
	\caption{(Color online) Same as bottom panel of Fig.~\ref{fig2}, but for $L=15$.}
	\label{fig2add}
\end{figure}

In confirming these results for larger values of $L$, such as $L=15$,
we have identified in the latter case 8 unstable pairs (say,
for $\omega_b < 2$). Among these,
5 pairs are found to show an increasing tendency, 2 to be decreasing, while
1 remains nearly invariant over our variation of $\omega_b$.
Once again, no new unstable eigenmodes arise over the parametric
scan --see Fig.~\ref{fig2add}.

When $\omega_b$ is small, the Floquet exponents $\sigma$ approach finite
values --see the bottom panel of Fig.~\ref{fig2}-- and we have checked that
their unstable eigenmodes approach a limiting profile (data not shown).
  As computational power used for this problem is further enhanced,
  the  Peregrine limit of $\omega_b \rightarrow 0$ of the KMb
  will become progressively more accessible.
  Nevertheless, it is important to appreciate
that while going to lower frequencies, the
periods of the breathers start growing to high values,
and hence performance of the relevant computation becomes
exceptionally demanding. This is not so much true at the
level of the direct time integration,
but most notably at that of the solution of the variational equations
for the stability analysis
which constitute a system of size $N^2 \times N^2$ (where $N$
is the number of grid points used). While this is usually
done for {discrete} breathers~\cite{flach}, typically the number of
nodes can be restricted to $N \approx 100$ or so. However, in our
case this being a {\it continuum} computation, the number of
grid points to resolve the solution --even more notably so
because of the space-time decay of a Peregrine soliton-- is particularly
large.
More specifically, here we use $N=40L$ (typically $L=10$ is
  implemented), in accordance with the discretization parameter
  $\Delta x=0.05$ of the numerical integrator.

While we cannot fully extrapolate our results to the Peregrine limit,
we do note that the trend of the instabilities observed is such
that they only stem from the background and not from the localized
core of the solution, a feature suggestive towards the relevant limit.

\subsection{Direct Numerical Simulations}

To elucidate the role of the above found instabilities %found above
on the dynamics, we have performed a series of direct numerical simulations
for the NLS Eq.~(\ref{eq:nls}). The simulations concern the evolution
of the KMb, being set as initial condition (at $t=0$, as $u_{\mbox{\tiny KM}}(x,t=0)$),
perturbed along the direction of different unstable eigenmodes $\xi$;
{that is, the initial conditions used are of the form
$u(x,t=0)=u_{\mbox{\tiny KM}}(x,0)+\delta \xi(x,0)$, with $\delta=10^{-3}$}.
Simulations have been performed for $L=10$ and $\omega_b = 15$, {although, for the sake of clarity, the
figures only show the interval $[-5,5]$. Notice that for the value of $\wb$ taken in the simulations, if
$\delta=10^{-3}$ then $|u(\pm L,0)-1|\sim10^{-5}$, so boundary effects are negligible.}

%%%%%%%%%%%%%%%%%%%%%%%%%%%%%%%%%%%%%%%%%555
\begin{figure}[h]
	\begin{center}
		\begin{tabular}{cc}
			\includegraphics[width=.34\textwidth]{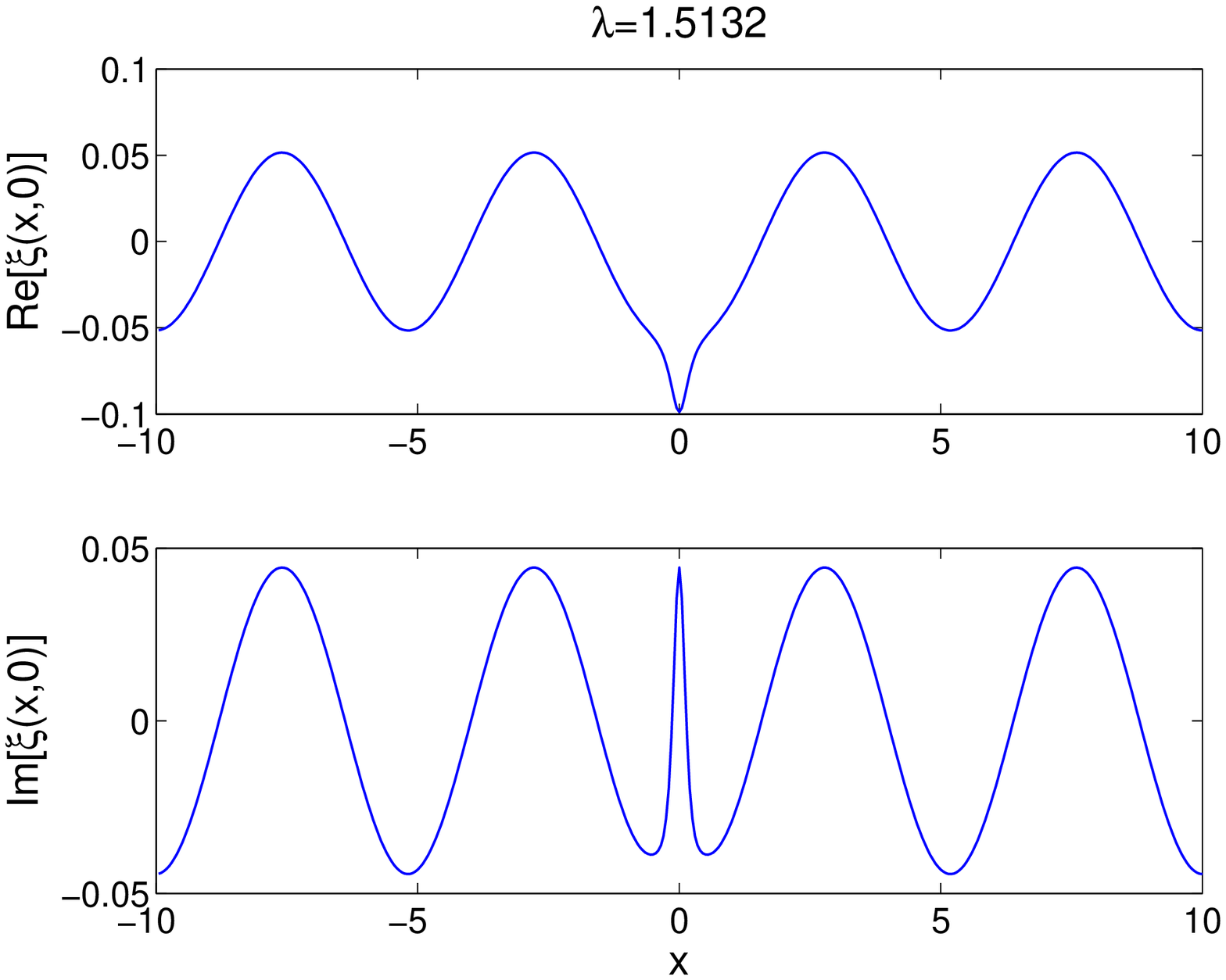} \\ %&
			\includegraphics[width=.33\textwidth]{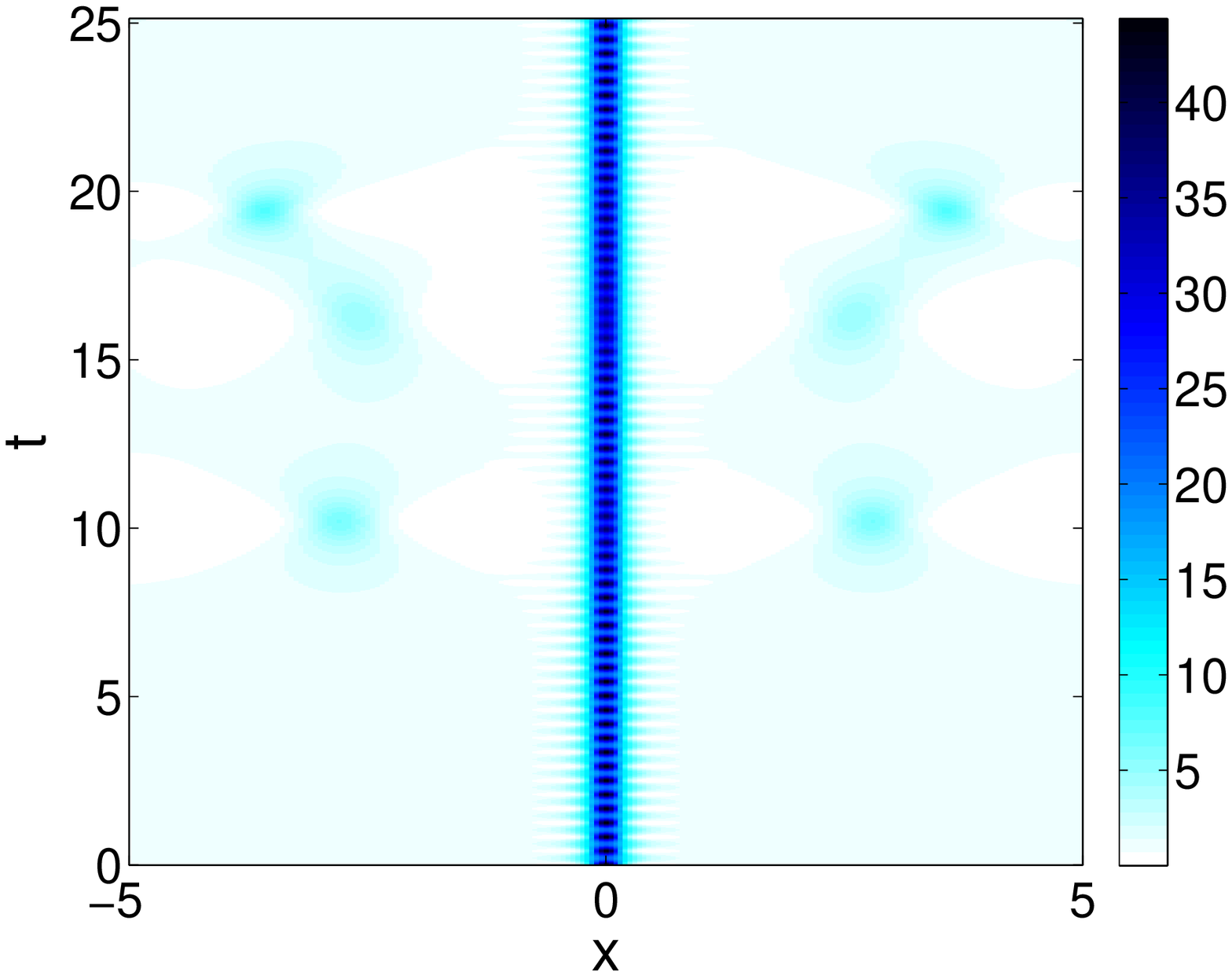} \\
			\includegraphics[width=.33\textwidth]{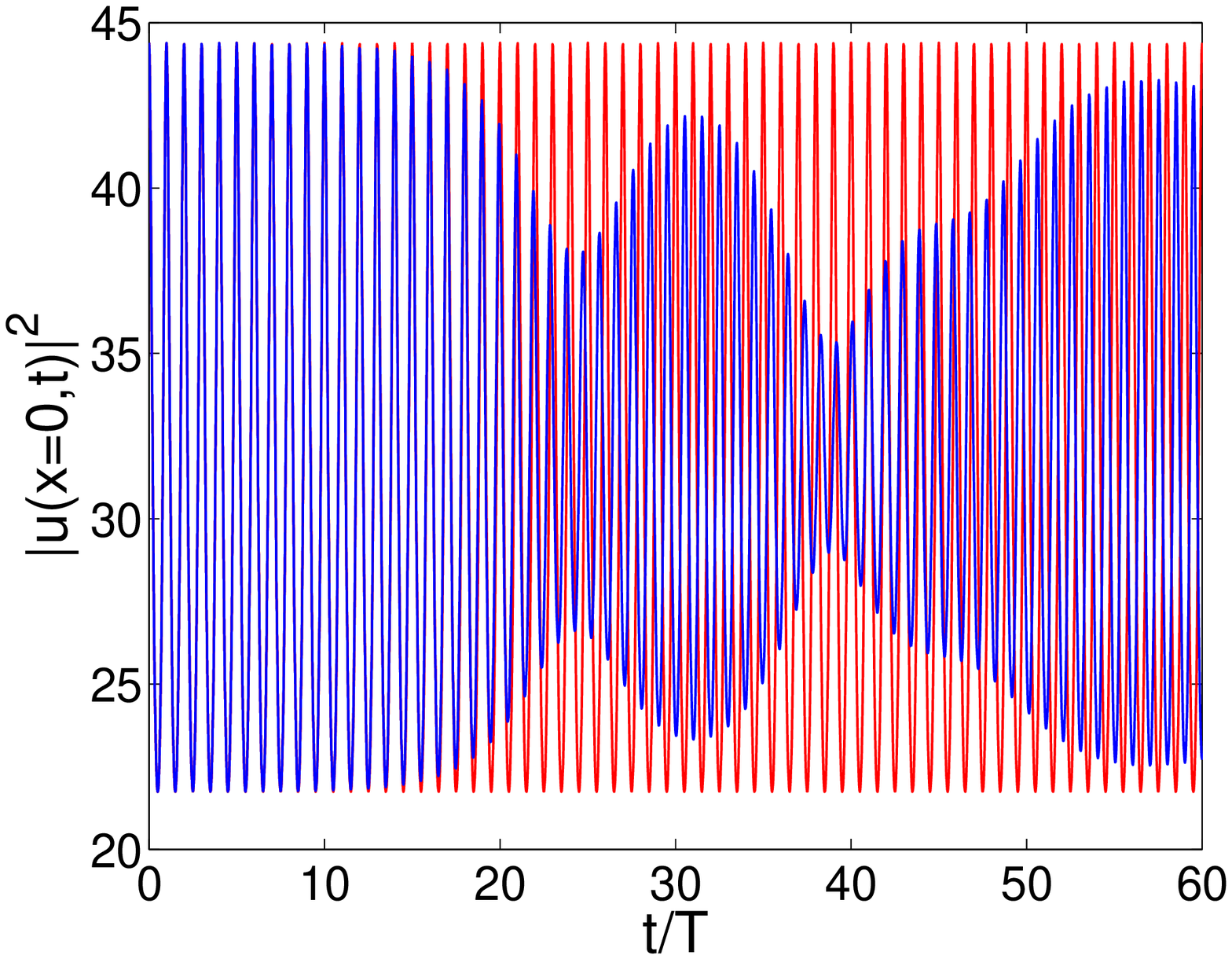}
		\end{tabular}
	\end{center}
	\caption{(Color online) KMb with $L=10$ and $\omega_b=15$.
		Top two panels: The spatial profiles, $\mathrm{Re}[\xi(x,0)]$ (first)
		and $\mathrm{Im}[\xi(x,0)]$ (second),
		of the unstable even eigenmode corresponding to $\lambda=1.5132$. %Middle
		Third panel: Contour plot of the evolution of the density $|u(x,t)|^2$, corresponding to a KMb initial
		condition (at $t=0$), perturbed along the direction of the unstable eigenmode
		portrayed in the top two panels. Bottom panel: comparison of the evolution of
		the center density (at $x=0$) of the breather [dark grey (blue) line], associated with the
		density evolution of the middle panel, against the respective evolution of
		the exact analytical KMb solution [light grey (red) line].}
	\label{fig3}
\end{figure}
%%%%%%%%%%%%%%%%%%%%%%%%%%%%%%%%%%%%%%%%%%%%%%%%%%%%%%%%

\begin{figure}[tbp]
	\begin{center}
		\begin{tabular}{cc}
			\includegraphics[width=.34\textwidth]{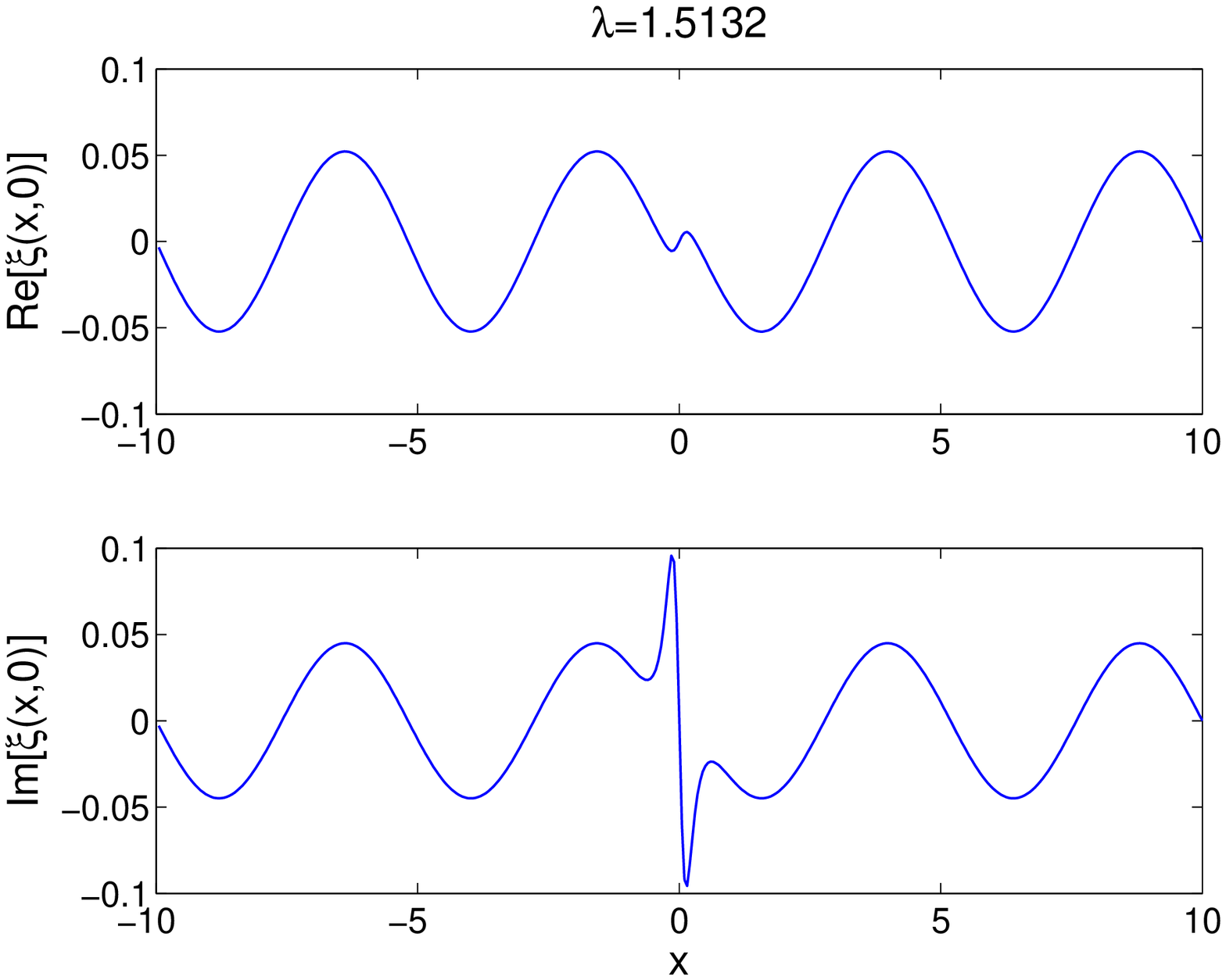} \\
			\includegraphics[width=.33\textwidth]{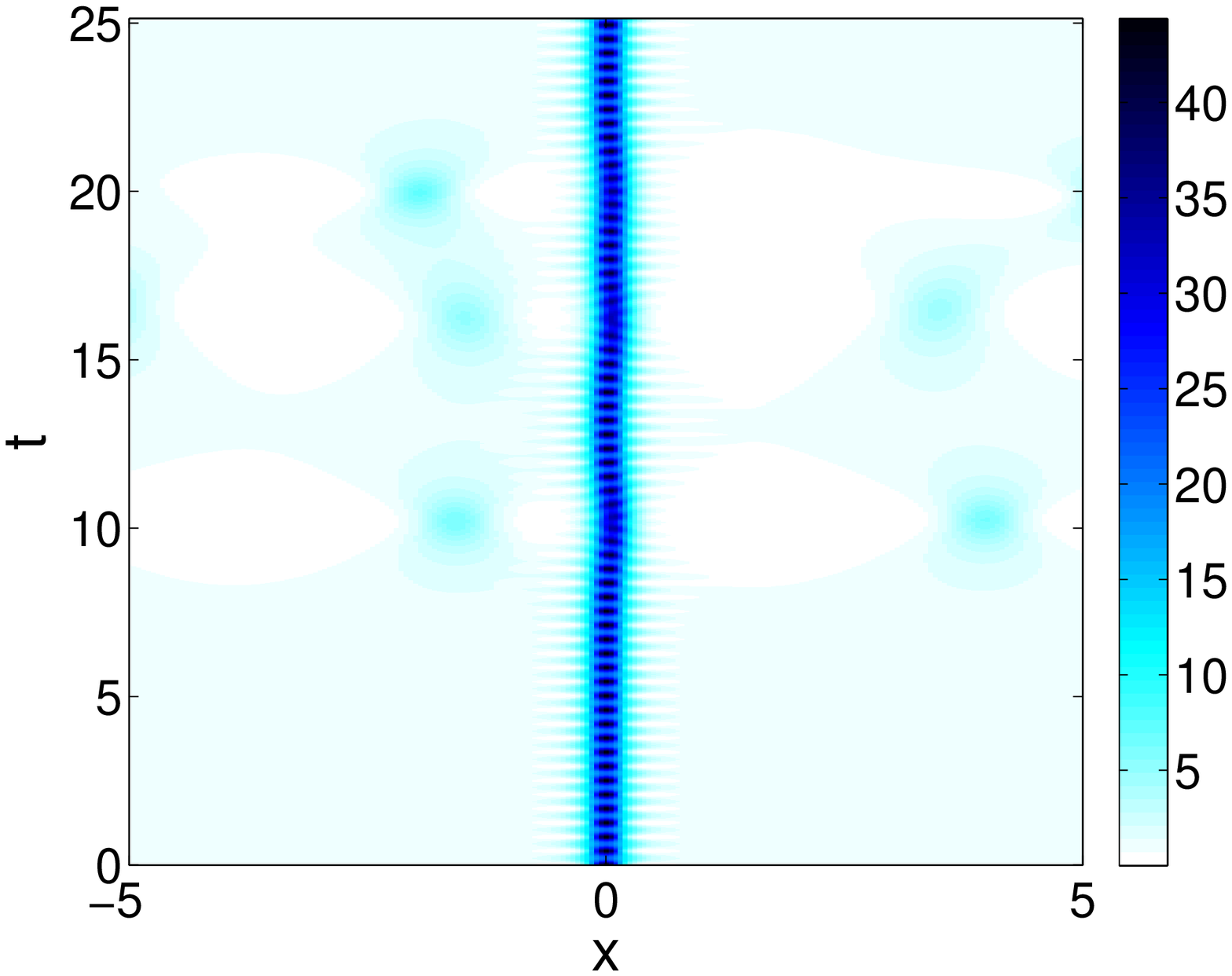}
		\end{tabular}
	\end{center}
	\caption{(Color online) Same as the three top panels of Fig.~\ref{fig3}, but
		for the unstable odd eigenmode, corresponding to $\lambda=1.5132$.
		Here, it is observed that, due to the instability, the KMb performs a small-amplitude
		oscillation around its original location.}
	\label{fig4}
\end{figure}

Before presenting our numerical findings, it is worth discussing at first
the structure of the eigenmodes associated with the unstable and neutral eigenvalues.
Generally, we have found that to each unstable eigenvalue corresponds a pair
of degenerate, parity symmetric eigenmodes, with
each one possessing even or odd symmetry. Both eigenmodes feature,
asymptotically (for large $|x|$) an extended -- periodic -- structure, and
are characterized by the presence of a localized peak (or dip) at the location of the KMb
(at $x=0$); this peak is either even or odd for the respective even and odd eigenmode.
Typical spatial profiles of the real and imaginary parts of even and odd eigenmodes,
corresponding to $\lambda=1.5132$, are depicted
in the two top panels of Figs.~\ref{fig3} and \ref{fig4}, respectively.
For the limiting case of the neutral exponent with $\lambda=1$,
the eigenmodes become localized in the vicinity of the KMb, tending asymptotically to zero.

We now proceed by investigating the effect of the above types of eigenmodes on the
dynamics of the KMb. We observe that a perturbation along spatially even modes,
such as the one pertinent to the unstable eigenvalue $\lambda=1.5132$
(cf. top two panels of Fig.~\ref{fig3}), leads to the evolution depicted
in the third and the bottom panels of Fig.~\ref{fig3}, where a modulation of the breather,
as well as a change of its frequency, are observed. Similar dynamics, although accompanied by a small-amplitude oscillation in the position
of the breather, is also observed when the perturbation is along an odd mode eigendirection, such as the
one corresponding to $\lambda=1.5132$ (cf. top two panels of
Fig.~\ref{fig4}). Notice that, in this case, the even KMb is perturbed by an odd eigenmode,
which results in the emergence of this small-amplitude oscillation.
The latter, is evident in the density evolution plot, shown in the third panel of Fig.~\ref{fig4}.

%%%%%%%%%%%%%%%%%%%%%%%%%%%%%%%%%%%%%%%%%555
%%%%%%%%%%%%%%%%%%%%%%%%%%%%%%%%%%%%%%%%%%%%%%%
\begin{figure}[tbp]
	\begin{center}
		\begin{tabular}{cc}
			\includegraphics[width=.34\textwidth]{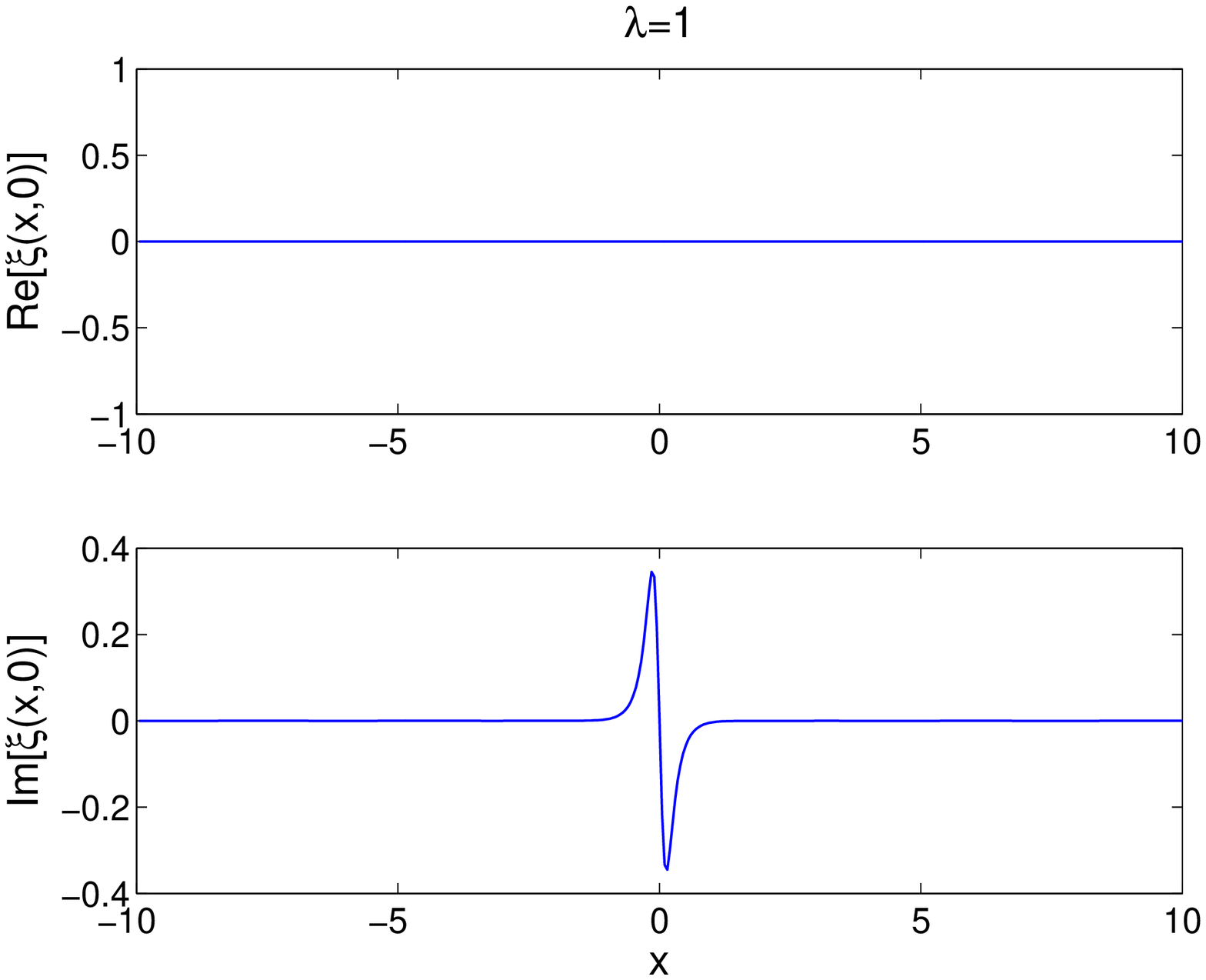} \\ %&
			\includegraphics[width=.335\textwidth]{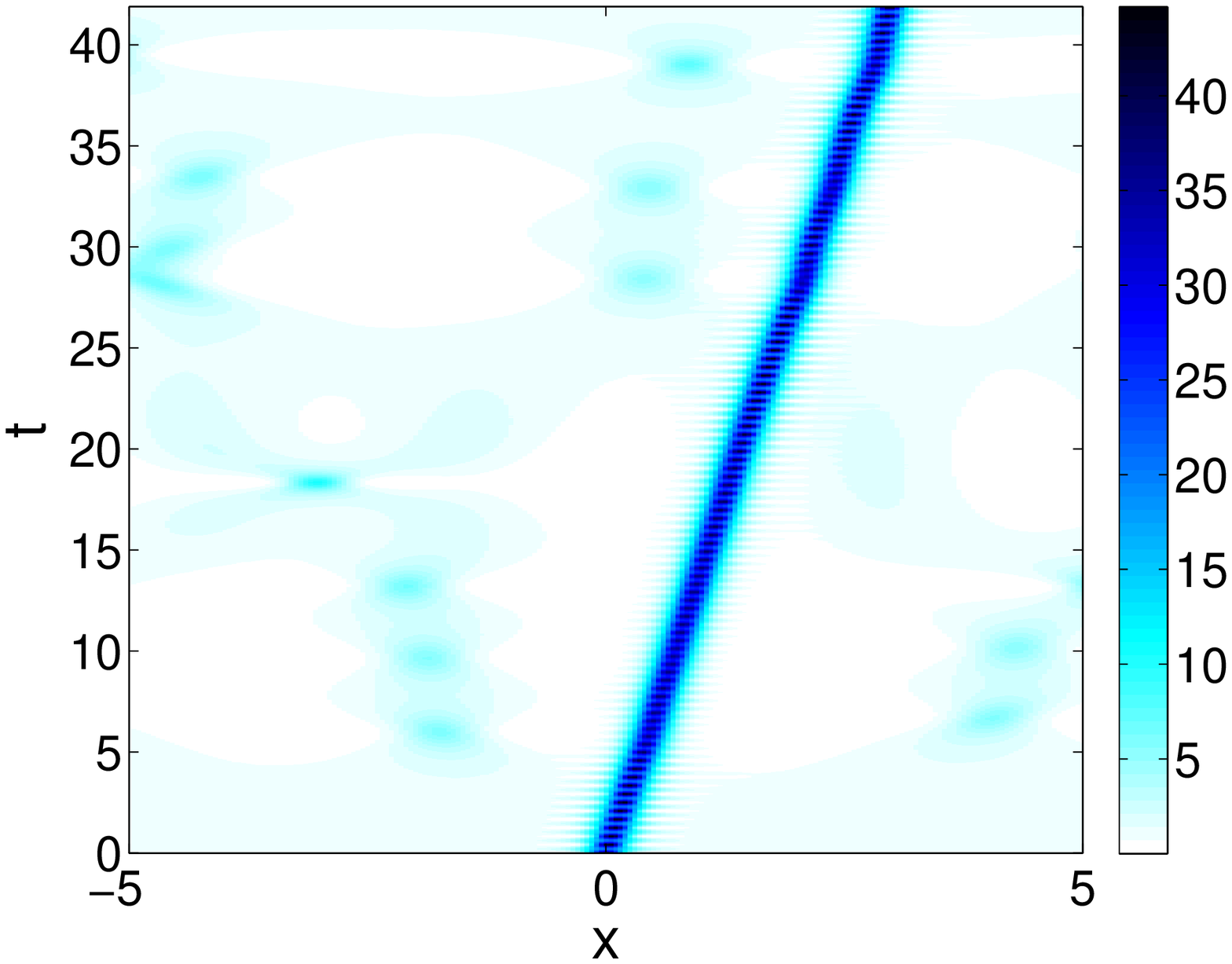} \\
			\includegraphics[width=.335\textwidth]{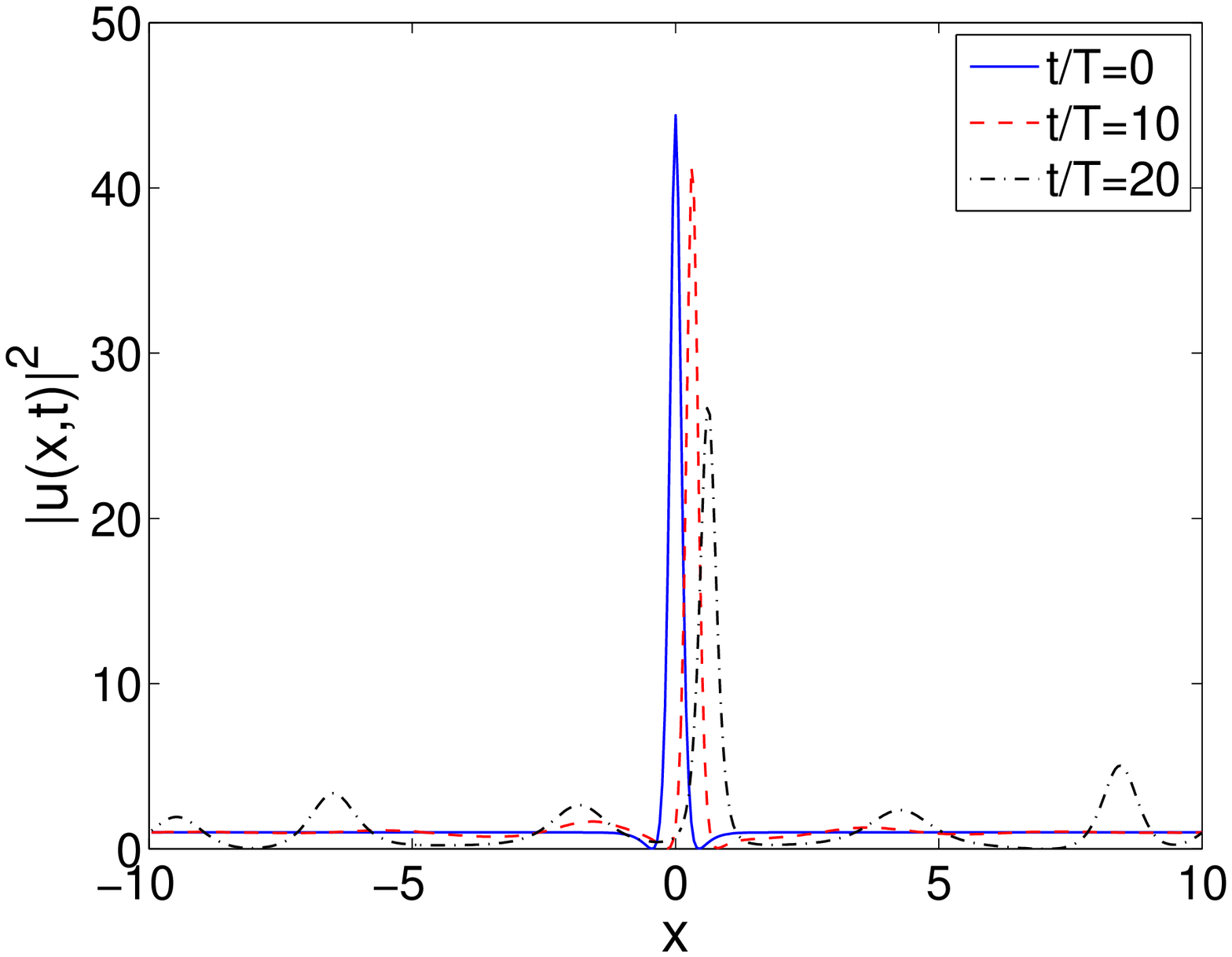} \\
		\end{tabular}
	\end{center}
	\caption{(Color online) Same as Fig.~\ref{fig4}, but for
	  the translational mode, corresponding to $\lambda=1$.
          The bottom panel shows the KMb profile at $t=0$ and when the breather has oscillated 10 and 20 periods. Notice the deformation of the background and the change of the maximum breather amplitude.}
	\label{fig5}
\end{figure}
%%%%%%%%%%%%%%%%%%%%%%%%%%%%%%%%%%%%%%%%%%%%%%%%%%55

If the KMb is perturbed along the translational mode, it becomes mobile, as shown in Fig.~\ref{fig5}. However, its evolution cannot be considered as following the typical traveling wave profile (\ref{eq:gali}): the maximum amplitude of the breather changes in an oscillatory fashion when moving along the domain  because of the interaction of the breather with the deformed background.

Notice that the dynamics of the KMb in all of the above cases
takes place on top of
the modulationally unstable background, as is evident in all
space-time contour plots of Figs.~\ref{fig3}, \ref{fig4}, and \ref{fig5}.
Thus, while the KMb does not itself get destroyed, the underlying
background gets distorted (as a result of the instability) during
the evolution.

\section{Conclusions and Future Challenges}
In the present work, we took a step towards exploring the spectral stability of
rogue waves.
In particular, we argued that given that these extreme events are limiting cases
of other periodic (in the evolution variable) structures, it is possible to perform at first
a Floquet stability analysis of the latter, and then try to
{\it a posteriori} obtain suitable clues for the limiting case.
This way, we focused on the spectral stability of the Kuznetsov-Ma breather (KMb), which --in
the limit of infinite period-- reduces to the Peregrine soliton (PS).

Our conclusions are two-fold.
On the one hand, we discovered that
the unstable modes commonly have
a stronger instability growth rate in the presence
of the KMb rather than in its absence, a conclusion which
seems to be persisting as a trend as the frequency
is decreasing towards the PS limit $\omega_b \approx 0$. In other words,
this finding suggests that as the frequency decreases, the structure
is more unstable than the background
itself.
This may be interpreted as stemming from the additional energy and the
``seeding'' of the instability of the background induced by the presence
of the KMb.
On the other hand, we observed
the absence of any additional
point spectrum (localized eigendirections), under the continuation of the KMb
towards the Peregrine limit. This absence of point spectrum, intriguingly
suggests the following:
if some additional terms (linear/nonlinear) were incorporated in the original NLS model,
such that {\it a partial stabilization of the background} was achieved, then the PS state would
be more robust and, hence, more easily observable.
Indeed, such a partial stabilization of the PS seems to
to be the case in water waves, and similar experimentally tractable examples \cite{k2a}-\cite{cap}.
Moreover, the perturbations that lead to dynamical instability, as shown
in Figs.~~\ref{fig3}--\ref{fig5} may modify the frequency of the KMb
or result in its mobility, but {\it are not detrimental} to the localized
component of the structure per se.
Based on the above arguments, we believe
that, indeed, it is the above PS stabilization scenario which is relevant to the observability
of the, otherwise unstable, rogue waves.

The present study, opens a number of paths for future exploration.
Given the availability of the KMb in an explicit analytical
form, while technically challenging, it is worthwhile to potentially
ponder on the analytical tractability of the Floquet multiplier
problem (especially in the light of the integrability of the 1D
NLS model).
Another interesting and relevant direction, is
to explore the possibility of generalizing KMb solutions
(and thereafter, their stability properties), to
other models, such as
extended NLS equations incorporating higher order effects~\cite{devine,themis}.
Finally, exploring whether
Peregrine soliton solutions (and by extension, Kuznetsov-Ma breathers)
may exist in higher-dimensional systems, is an intriguing topic
in its own right, that merits further investigation. Some of these
aspects are currently under consideration, and will be reported in future
work.

\appendix
\setcounter{section}{0}
\label{apFT}
\section{On the
%Elements of
Floquet stability theory}

In this Appendix, we provide brief information on the definition of the
monodromy matrix $\mathcal{M}$ defined in Eq.~(\ref{MM}), and its relevance to the
stability problem of time-periodic solutions as the Kuznetsov-Ma breather (KMb) of Eq~(\ref{eq:nls}).

The linearized equation Eq.~(\ref{eq:perturb}), can be written in the form of
a system of first-order, linear evolution equations as follows:
\begin{equation}
\label{Ap1}
\frac{d}{dt}V=\left[\begin{matrix}
0&\mathcal{L}_{-} \\-\mathcal{L}_{+}&0 \end{matrix}\right]V=\mathcal{J}(t)V,
\end{equation}
where $V=\left[\mathrm{Re}(\xi(x, t)), \mathrm{Im}(\xi(x, t))\right]^T$.
The linear differential operators in the matrix $\mathcal{J}(t)$ of  Eq.~(\ref{Ap1}),
are defined as
$\mathcal{L}_{-}=-\frac{1}{2}\partial^2_x-2|u_{\mbox{\tiny KM}}|^2+1+u_{\mbox{\tiny KM}}^2$, and
$\mathcal{L}_{+}=-\frac{1}{2}\partial^2_x-2|u_{\mbox{\tiny KM}}|^2+1-u_{\mbox{\tiny KM}}^2$.
Due to the
presence of the KMb solution $u_{\mbox{\tiny KM}}(x,t)$,
the matrix $\mathcal{J}(t)$, is time-periodic with period  $T=2\pi/\wb$, i.e., $\mathcal{J}(t+T)=\mathcal{J}(t)$.

By numerical integration of (\ref{Ap1}) as described in Sec.~\ref{SecA}, we find its numerical (fundamental) solution $V(t)$, satisfying the relation
\begin{eqnarray}
\label{Ap2}	
V(t+T)=V(t)V^{-1}(0)V(T).
\end{eqnarray}
Equivalently to Eq.~(\ref{MM}), the monodromy matrix is defined by Eq.~(\ref{Ap2}),
as $\mathcal{M}=V^{-1}(0)V(T)$. On the other hand, Floquet's theorem \cite{Pkh,Li}
allows to reduce the  fundamental solution of Eq.~(\ref{Ap1}) in the form
$V(t)=\Phi (t)\exp(t R)$, where $R$ is a constant matrix, and $\Phi(t)$ is periodic.
Due to periodicity, it holds that
\begin{eqnarray}
\label{Ap3}
V(t+T)&=&\Phi (t+T)\exp[(t+T)R]\nonumber\\
&=&V(t)\exp(T R).
\end{eqnarray}
Then, comparing Eq.~(\ref{Ap2}) to Eq.~(\ref{Ap3}), we may relate the
monodromy matrix $\mathcal{M}$ with the matrix $R$, by the formula
\begin{eqnarray}
\label{Ap4}	
\exp(T R)=V^{-1}(0)V(T)=\mathcal{M}.
\end{eqnarray}
It is evident from Eq.~(\ref{Ap4}), that the eigenvalues of the monodromy matrix $\mathcal{M}$,
the Floquet multipliers (FMs), determine the stability of the KMb (as they are associated through  Eq.~(\ref{Ap4}), to the eigenvalues of $R$, the corresponding Floquet exponents).
The possibilities for FMs, and the induced instabilities, are as discussed in Sec.~\ref{SecA}.

\vspace{.5cm}

{\bf Acknowledgments.} P.G.K., D.J.F. and N.I.K. acknowledge that this work was made
possible by NPRP grant {\#} [8-764-160] from Qatar National Research Fund (a member of Qatar Foundation).
The findings achieved herein are solely the responsibility of the authors.
M.H. acknowledges the support of the ANR project BoND (ANR-13-BS01-0009-01) and the Labex ACTION program (ANR-11-LABX-01-01). J.C.-M. thanks financial support from MAT2016-79866-R project (AEI/FEDER, UE).

\end{document}